\documentstyle[pre,multicol,aps,psfig]{revtex}
\begin{document}
\draft
\title{Dissipative Dynamics of Collisionless Nonlinear Alfv\'en Wave Trains}
\author{M. V. Medvedev$^{1,}$\cite{email,mm}
, P. H. Diamond$^{1,}$\cite{pd}
, V. I. Shevchenko$^2$
, and V. L. Galinsky$^3$ }
\address{$^1$ Physics Department, University of California at San Diego,
La Jolla, California 92093-0319}
\address{$^2$ Electrical \& Computer Engineering Department,
University of California at San Diego, La Jolla, California 92093-0407}
\address{$^3$ Scripps Institution of Oceanography,
University of California at San Diego, La Jolla, California 92093-0210}
\maketitle

\begin{abstract}
The nonlinear dynamics of collisionless Alfv\`en trains, including resonant particle 
effects is studied using the kinetic nonlinear Schr\"odinger (KNLS) equation model.
Numerical solutions of the KNLS reveal the dynamics of Alfv\'en waves to be sensitive
to  the sense of polarization as well as the angle of propagation with
respect to the ambient magnetic field. The combined effects of both wave nonlinearity
and Landau damping result in the evolutionary formation of {\em stationary} 
S- and arc-polarized directional and rotational discontinuities. These waveforms
are freqently observed in the interplanetary plasma.
\end{abstract}

\begin{multicols}{2}
Numerous satellite observations of magnetic activity in the solar wind
have exhibited the nonlinear nature of MHD waves \cite{NLwaves1,NLwaves2}.
Recent observations indicate the existence
of directional (DD) and rotational (RD) discontinuities, i.e.
regions of rapid phase jumps where the amplitude also varies 
\cite{NLwaves1,DD}, which are thought to be a result of the nonlinear development 
and evolution of MHD waves. Several types of RD/DDs
which might be distinguished by their phase portraits
have been observed. There are
(i) discontinuities of the {\em ``S-type''}, at which the magnetic field 
vector rotates
first through an angle less than or approaching $90^\circ$ in one direction,
followed by rotation in the opposite direction through an angle larger 
than $180^\circ$ (typically, $180^\circ<\Delta\phi\le270^\circ$) 
\cite{DD,S-type}, and (ii) {\em arc-polarized}
discontinuities, where the magnetic field vector rotates along an arc through
an angle less than $180^\circ$ \cite{NLwaves1,arc}.
At DDs, the fast phase jump is accompanied by
moderate amplitude modulation ($\delta B\sim B$). At RDs,
the amplitude modulation is small or negligible
($\delta B\ll B$).

The envelope dynamics of nonlinear Alfv\'en waves at small-$\beta$
are thought to be governed by the derivative nonlinear Schr\"odinger (DNLS) 
equation, which describes parametric coupling with acoustic modes 
\cite{Cohen}. The theory of nondissipative  Alfv\'en waves governed by the 
conservative DNLS equation predicts nonlinear wave steepening and 
formation of waveforms with steep fronts. 
Thus, spiky, many-soliton structures are
emitted from the steep edge. 
It was shown that the nonlinear wave relaxes
to a shock train and constant-$B$  RDs where the field 
rotates through exactly $180^\circ$ \cite{Cohen,train}, when the linear 
damping due to finite plasma conductivity is taken into account.
Inspite of this, the DNLS theory was unable to explain the existence
and dynamics of both (i) the S-polarized DDs and 
RDs and (ii) arc-polarized RDs,
with rotation of less than $180^\circ$. 
It is believed (and confirmed by recent particle code simulations
\cite{Vasquez1,Vasquez2,Vasquez3}) 
that the dynamics of Alfv\'en waves in the $\beta\sim1$, isothermal solar 
wind plasma are intrinsically {\em dissipative}, on account of 
Landau damping of ion-acoustic oscillations 
\cite{Rogister-Mjolhus-Spangler,us}.
We should comment here, that particle code simulations (e.g, Refs.  
\cite{Vasquez1,Vasquez2,Vasquez3}) model ``microscopic'' behavior of plasma
particle motions, thus may be refered to as ``numerical experiments''. A numerical
solution of a ``macroscopic'' evolution equation is a complementary way which
allows to get a theoretical insight into the underlying physics and theoretically
explain the observed (experimental) data.

The dynamics of magnetic fields in the solar wind has been extensively investigated
using different analytical approaches. Various (e.g., beat, modulational,
decay) instabilities of Alfv\'en waves were shown \cite{Hollweig94} to be sensitive 
to the wave polarization and the value of plasma $\beta$. The turbulence-based
linear model which describes the radial evolution of magnetic fluctuations 
in the solar wind was successfully developed \cite{Zank-Mattheus=Oughton-Mattheus}
taking into account the effects of advection and solar wind expansion, along with mixing
effects. The simple nonlinear noisy-KNLS model of turbulence was proposed and 
investigated in \cite{usDIA}. Arc-polarized waves, which had been first discussed 
in \cite{Barnes-Hollweg}, were explained \cite{Vasquez-Hollweg}
via coupling of obliquely propagating 
circular Alfv\'en waves and a driven fast/slow wave. Some damping is necessary
in this case in order to (consistently) select the arc-type solutions.

In this paper, we use a recently developed \cite{us} analytical model of the 
kinetic DNLS (KNLS) equation to investigate the influence of Landau damping 
on the (strongly) nonlinear dynamics of Alfv\'en waves. 
The main claim of this paper is that {\em all} the discontinuous wave structures
discussed above are {\em distinct solutions} of the same {\em simple} analytical
model for different initial conditions, e.g. initial wave polarization
and wave propagation angle. One should note that the term which describes 
the resonant particle effect is usually integral 
(nonlocal) in nature, on account of the finite ion transit time through the envelope
modulation of an Alfv\'en train. Thus, the envelope evolution equation 
obtained is a nonlocal,
integro-differential equation which is not amenable to 
{\em analytical} solution. We should comment here a case which
extends the traditional paradigm of shock waveforms. There are two known types 
of shocks,
namely (i) {\em collisional {\em (hydrodynamic)} shocks}, in which 
nonlinear steepening is limited by {\em collisional} (viscous) dissipation,
which sinks energy from small scales and
(ii) {\em collisionless shocks} (common in astrophysical plasma),
in which nonlinear steepening is limited by {\em dispersion}, resulting in the
formation of
soliton-type structures with energy content in high-$k$ harmonics. We add a new class 
of shock, namely (iii) {\em dissipative structures} (which also can be referred 
to as {\em collisionless dissipative shocks}), for which nonlinear steepening 
is limited by {\em collisionless} (scale independent, i.e., acting on all scales)
damping. They emerge only from quasi-parallel, (nearly)
linearly polarized waves. 
Of course, to obtain familiar shock-like wave forms, the cyclotron
damping at large-$k$ must be incorporated, as usual for
collisionless shocks. 

The KNLS equation may be written \cite{Rogister-Mjolhus-Spangler,us} as 
\begin{eqnarray}
\frac{\partial b}{\partial t}&+&\frac{v_A}{2}\frac{\partial}{\partial z}
\Bigl(\left\{M_1\left(|b|^2-\langle|b|^2\rangle\right)
\right.\Bigr.\nonumber\\ & & { }\left.\left. 
+M_2\widehat{\cal H}\left[|b|^2-\langle|b|^2\rangle\right]\right\}b\right)
+ i\frac{v_A^2}{2\Omega_i}\frac{\partial^2b}{\partial z^2}=0 ,
\label{mainKNLS}
\end{eqnarray}
where $b=(b_x+ib_y)/B_0$ is the wave magnetic field,
$v_A$ and $\Omega_i$ are the Alfv\'en speed and proton ion-cyclotron 
frequency, $\langle...\rangle$ means average over space and fast (Alfv\'enic) time. 
Here the constants $M_1$ and $M_2$ depend on $\beta$ only
(see Ref.\ \cite{us} for details) and $\widehat{\cal H}$ is the  integral 
Hilbert operator $\widehat{\cal H}[f](z)={\pi}^{-1}
{\textrm{--}\!\!\!\!\int_{-\infty}^\infty}
{(z'-z)}^{-1}f(z') {\rm d}z'$. This equation was solved
for periodic boundary conditions using a predictor-corrector scheme and
a fast Fourier transform, 1024 harmonics and spatial points were taken.
The dimensionless spatial coordinate and time were introduced
respectively as $\zeta=z/\bar z$ and $\tau=t_e/\bar t$,
where $\bar z=50c/\omega_p$ and $\bar t=200/\Omega_i$.
For $\beta=0$, kinetics do not impact the wave dynamics \cite{us}, so
Eq.\ (\ref{mainKNLS}) reduces to the familiar DNLS equation
($M_1=0.5, M_2=0$).
The DNLS is integrable and has an exact (soliton) solution.
The test run has shown an excellent agreement
with the analytical solution during the time of computation 
(up to $\tau=40$, i.e. 8,000 cyclotron periods).

High-amplitude magnetic perturbations in plasmas typically evolve
from small-amplitude (linear) ones. Thus the most general approach is to 
examine the nonlinear evolution of finite-amplitude periodic waves of 
different polarizations. The initial wave profiles are given by two initially
excited Fourier harmonics, $b_k$. For linear polarizations, we pick  $b_k$'s equal
$b_{-1}=b_1=1$, all others are zeroes, for circular polarizations
we pick $b_{-2}=b_{-1}=1$, for elliptical polarizations, we  pick
$b_{-1}=1.1,~ b_{1}=0.9$. Thus, the waves are left-hand polarized.
Results of a reference run for $\beta=0$, with amplitude modulated 
linear polarization (the circular polarization and oblique propagation cases 
look similar) are shown in Fig.\ \ref{fig:lincomp-prl}a.
The wave exhibits  the nonlinear steepening phase of a front
at early imes ($\tau\sim2$). Dispersion further limits steepening
and produces (at times $\tau\sim5$) small-scale,
oscillatory, circularly polarized wave structures (even for 
initial linear polarization). 
Significant high harmonic energy content is generated.
At later times, $\tau\sim40$ (not shown), nonlinear processes result in 
a wave magnetic field which is completely irregular, 
indicating strong, large-amplitude Alfv\'enic turbulence.

For $\beta\not=0$, we first compare the waveform profiles and harmonic 
energy spectra obtained from the KNLS with the previous case. From now on,
$\beta=1$ and $T_e=T_i$ ($M_1=0.75, M_2=-0.83$)
unless stated otherwise. Fig.\ \ref{fig:lincomp-prl}b
depicts the time evolution of a parallel propagating wave with the same initial 
conditions. In contrast to the 
$\beta=0$ case, localized {\em quasi-stationary} structures are seen to 
form very rapidly, the formation time is $\tau_f\sim2$. 
The harmonic spectrum of the dissipative structures (also referred to as S-type DDs,
see below) is {\em narrow}, indicating that energy accumulates in low-$k$ harmonics, 
i.e.,  at large scales. It worth while to emphasize the quasi-stationary 
character of such waveforms. They preserve their shape for thousands 
cyclotron periods and thus may be indentified as ``structures''.  Meanwhile, 
the wave energy decays strongly, as seen from Fig.\ \ref{fig:energy-prl}.
Thus, the dissipative structures emerge via the competition of {\em nonlinear
steepening} of the wave and  scale invariant {\em collisionless damping}.
The fact that energy dissipates {\em in} the dissipative structures 
(and not somewhere inbetween) is readily
seen from the simple fact that $\widehat{\cal H}[const]\equiv0$.

Fig.\ \ref{fig:SDD-prl} is a snapshot of the dissipative structures at 
$\tau=15$ which emerged from a quasi-parallel, linearly polatized wave. 
It is seen that regions of significant field variations are
accompanied by fast phase rotation.
However, in the regions of negligibly varying $|b|$,  linear polarization
is preserved. The dissipative structures exhibit an easily distinguishable
{\em ``S-shaped''} phase portrait, namely that {\em at} the discontinuity
(solid path A-B-C), the magnetic field vector completes a rotation through 
$\pi$ radians. During the subsequent quescent 
region (path C-D), the magnetic field vector resides at the ``tip'' of  the left
arm indicating pure linear polarization. At the next discontinuity,
the vector returns to the initial position, similarly completing a $\pi$
radian rotation as shown by the dashed path.  Thus, the
KNLS dissipative structures have the requisite properties of
localized, {\em RDs/DDs}, as recently observed 
in the solar wind \cite{DD,S-type}. Note, however, these KNLS 
RDs are associated with the regions of {\em varying 
$|b|$}, unlike the conventional definition that $|b|=const$ across the 
RD. There is no sharp difference between an RD and a
(weak) DD. One may be transformed into another by changing
the initial wave polarization and propagation angle. Hence, we use both 
terms for the S-type KNLS discontinuity. We should also note the
remarkable similarity of hodographs obtained by solution of the KNLS
equation and from full numerical plasma simulations \cite{Vasquez1}.
Such KNLS discontinuities occur commonly and are not restricted
to $\beta$'s close to unity. These structures are quite evident in
a wide interval of $\beta$, of approximately 0.5-0.6 to
1.4-1.6. The dissipative structures still form at smaller $M_2$, however
the formation time increases when $M_2$ decreases. 

In contrast to the case of linear polarization,
circularly polarized, quasi-parallel waves evolve in a few $\tau$ to a 
single (almost purely) circularly polarized harmonic at the  lowest $k$
and do not form discontinuities. 
Energy decay (Fig.\ \ref{fig:energy-prl}) is negligible in the stationary state.

Fig.\ \ref{fig:piRD-prl} depicts a snapshot of the quasi-parallel,
initially elliptically polarized 
wave, an intermediate case between purely circular and linear polarizations.
Sudden phase jumps (by $\pi$ radians) which are localized at regions
of varying wave amplitude (typical of linear polarizations) are easily seen. 
However, these
discontinuities are not accompanied by wave amplitude discontinuities.
Thus, they are the $\Delta\phi=\pi$ {\em RDs}. Note, these 
discontinuities (which are the semi-circles in Fig.\ \ref{fig:piRD-prl}b) are
separated by extended regions of linear polarization. Energy dissipation 
(Fig.\ \ref{fig:energy-prl}) is weak, in comparison to the case of linear polarization.

Obliquely propagating waves are still described by the KNLS equation.
However, a new wave field which (formally) contains a perpendicular
projection component of the ambient magnetic field should be introduced.
Assuming the ambient field lies in $x$-$z$-plane, we write the new
field as $b=(b_x+B_0\sin{\Theta}+ib_y)/B_0$.
The nonlinear evolution of the linearly and highly elliptically polarized waves
is strongly sensitive to the angle between the polarization plane and the plane
defined by the ambient magnetic field vector and the direction of wave
propagation. This angle is set by initial conditions. When this angle is small,
the oscillating wave magnetic field has a longitudinal component
along the ambient field. Thus, we refer such waves to as {\em longitudinal}.
In the opposite case, the wave magnetic field oscillates (nearly)
perpendicularly
to the ambient field. Thus, such waves are called {\em transverse}. Note that this
classification scheme fails for circularly polarized waves, since a polarization plane 
cannot be defied in this case.

Fig.\ \ref{fig:arcRD-prl} shows a typical
(quasi-) stationary, {\em arc-polarized} discontinuity which evolved from 
an obliquely propagating ($\Theta\sim40^\circ$),
amplitude modulated, circularly polarized wave at $\tau=40$. 
The discontinuity is associated 
with minor (almost negligible) amplitude modulation.
The magnetic field vector makes a fast clockwise rotation through
less than $\pi$ radians (solid path A-B-C). The ends A and C are connected 
by a sector of circularly polarized wave packet (slow counterclockwise rotation
in the phase diagram, along the dashed, perfect arc C-D-A). Circular 
polarization is also indicated by the smoothly decreasing phase
outside the discontinuity (Fig.\ \ref{fig:arcRD-prl}a). 
Since $|b|^2\simeq const$ across the discontinuity
(as well as for a pure circularly polarized harmonic), it is nearly decoupled from
dissipation. Note the  remarkable similarity of this
solution of the KNLS equation to the structures detected in the solar wind
and observed in computer simulations \cite{arc,Vasquez3}.

The wave evolution of the
linearly polarized, obliquely propagating, transverse and longitudinal
waves differs drastically. The transverse waves
evolve very quickly (in a few $\tau$) to form a perfect arc-polarized 
RD. Energy dissipation is negligibly small in this
process. The longitudinal waves instead form
two S-type DDs propagating with different group 
velocities. Thus, they can merge and annihilate each other almost completely,
yielding a small-amplitude, residual magnetic perturbations. 

The sharp contrast between these three quasi-stationary solutions is a
direct consequence of the unique harmonic scaling of collisionless
(Landau) dissipation in the KNLS equation. It is crucial to understand that 
{\em collisionless damping enters at all $k$}, in contrast to hydrodynamic
systems where diffusion (viscosity) yields dissipation only at large $k$
(i.e. small scales, or steep gradient regions). However, higher-$k$ harmonics 
are strongly damped, which is typical of a phase-mixing process (i.e. smaller
scales mix faster). For quasi-parallel propagating waves, 
Landau damping enters symmetrically for $+k$ and $-k$ spectrum
components. It does not change the symmetry of a spectrum,
so that the initial helicity (set by initial spectrum symmetry)
is preserved. Since linear polarizations have
spectra symmetric upon $k\to-k$, they couple more strongly
to dissipation than circular polarizations do. Thus, S-polarized discontinuities,
which consist of predominantly two low-$k$ harmonics of nearly equal 
amplitude, emerge. For the circularly polarized wave, the (initial) spectrum is highly
asymmetric. Thus, such a wave
evolves to a single harmonic final state, which is, itself, a stationary 
(and exact) solution of the KNLS equation (i.e. it experiences no steepening
and minimal damping). No discontinuities emerge in this case.
For the oblique and quasi-perpendicular cases, there is asymmetry
between $b_x$ and $b_y$ components induced by the ambient magnetic field.
This allows the formation of a wave packet
with nearly constant $|b|^2$ (i.e., decoupled from dissipation). Such wave 
packets are the arc-polarized RDs with $\Delta\phi<\pi$.
We should emphasize the fact that since there is no 
characteristic dissipation scale in the system, the ultimate scale of the dissipative 
structures is set by {\em dispersion}, alone ({\em a l\`a} collisionless 
solitons and shock waves).
Accordingly, one can suggest that
(given initial equal populations of isotropically distributed circular and 
linear polarizations) quasi-parallel magnetic field fluctuations will
consist of predominantly circularly polarized waves and lower amplitude 
S-polarized KNLS DDs, while oblique perturbations
are predominantly arc-polarized discontinuities, separated by pieces of
oppositely circularly polarized waves. 

To conclude, the influence of the effect of Landau damping
was investigated in this paper. (A more complete study will be published 
elsewhere \cite{knls}.) A tractable analytic
evolution equation, the KNLS equation, was numerically solved to 
study nonlinear dynamics of finite-amplitude coherent Alfv\'enic trains
in a $\beta\sim1$, isothermal plasma, natural 
to the solar wind. Current studies shows that  
{\em all} the discontinuous wave structures observed in the solar wind 
are {\em distinct solutions} of the same {\em simple} analytical
KNLS model for different initial conditions, e.g. initial wave polarization
and wave propagation angle with no {\em a priori} assumptions or special 
initial conditions used.

We wish to thank B. T. Tsurutani, V. D. Shapiro, and S. K. Ride for useful 
discussions.
This work was supported by DoE Grant No.  DE-FG03-88ER53275, 
NASA Grants No.  NAGW-2418 and No. NAGW-5157,
and NSF Grant No. ATM-9396158  (with UC Irvine).

%
\end{multicols}
\onecolumn
\begin{figure}
\caption{ Wave profile evolution of a quasi-parallel, linearly polarized, 
sinusoidal wave initial condition for $\beta=0$ (a) and $\beta=1$ (b).}
\label{fig:lincomp-prl}
\end{figure}
\begin{figure}
\caption{Wave energy evolution for different initial conditions.}
\label{fig:energy-prl}
\end{figure}
\begin{figure}
\caption{S-polarized DD (quasi-parallel case),
(a) - amplitude and phase profiles,
(b) - hodograph.}
\label{fig:SDD-prl}
\end{figure}
\begin{figure}
\caption{Same as Fig.\ \ref{fig:SDD-prl} for the $\Delta\phi=\pi$ RD 
(quasi-parallel case).}
\label{fig:piRD-prl}
\end{figure}
\begin{figure}
\caption{Same as Fig.\ \ref{fig:SDD-prl} for the $\Delta\phi<\pi$ RD (oblique case).}
\label{fig:arcRD-prl}
\end{figure}
%
%
%
\newpage
\hspace*{-1in}\psfig{file=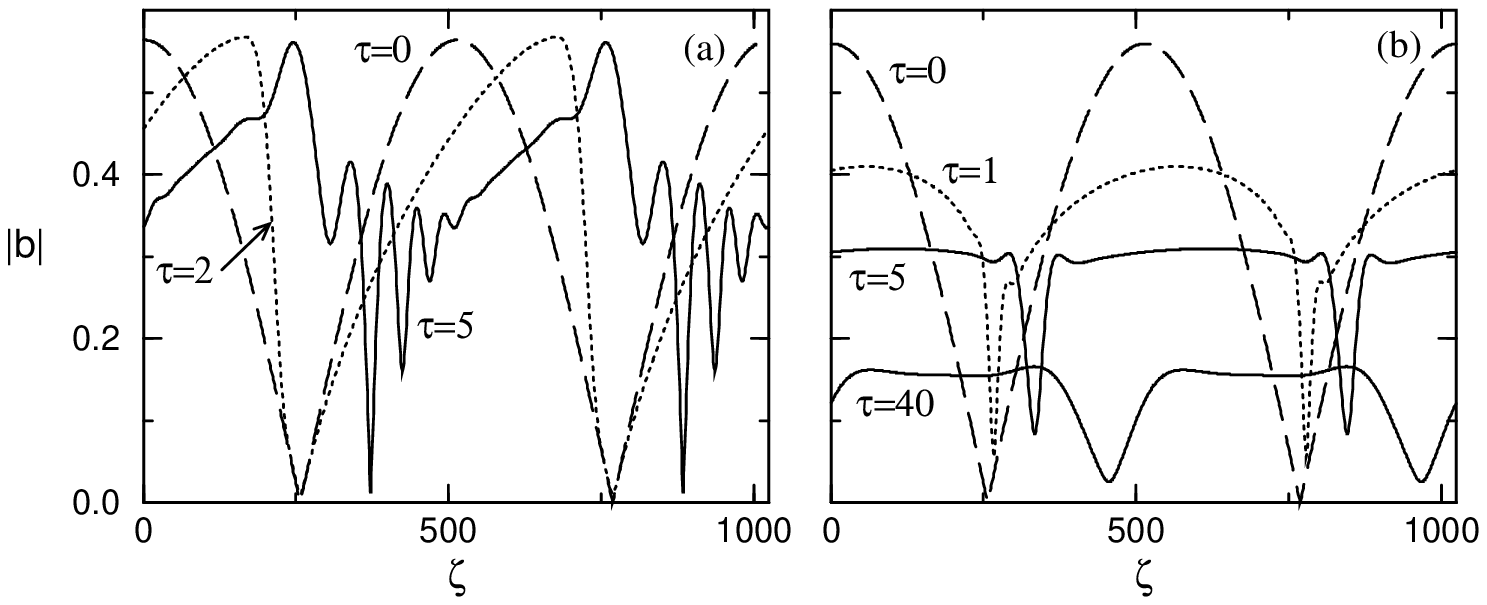}\vspace*{-8in} Fig.~1\vskip0.5in
\hspace*{-1in}\psfig{file=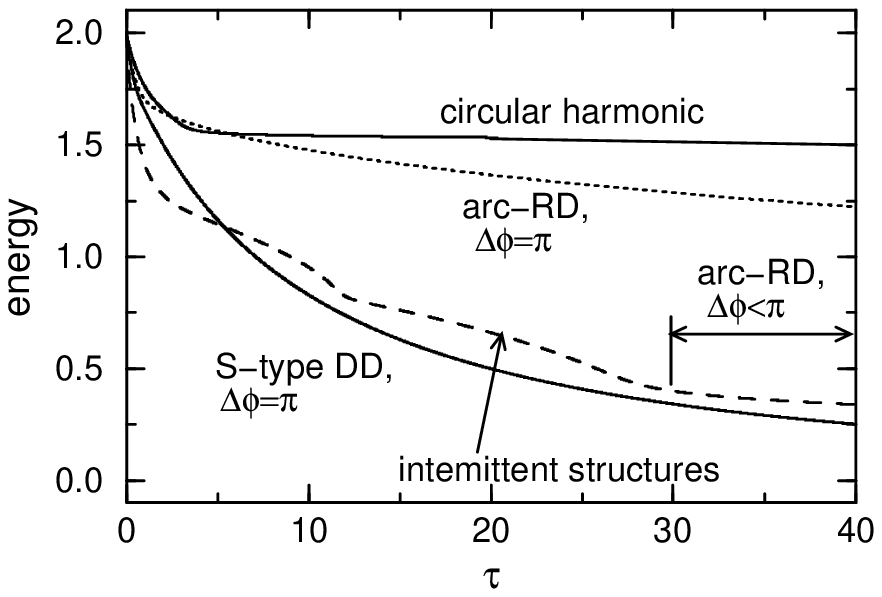}\vspace*{-8in} Fig.~2\vskip0.5in
\thispagestyle{empty}
\newpage
\hspace*{-1in}\psfig{file=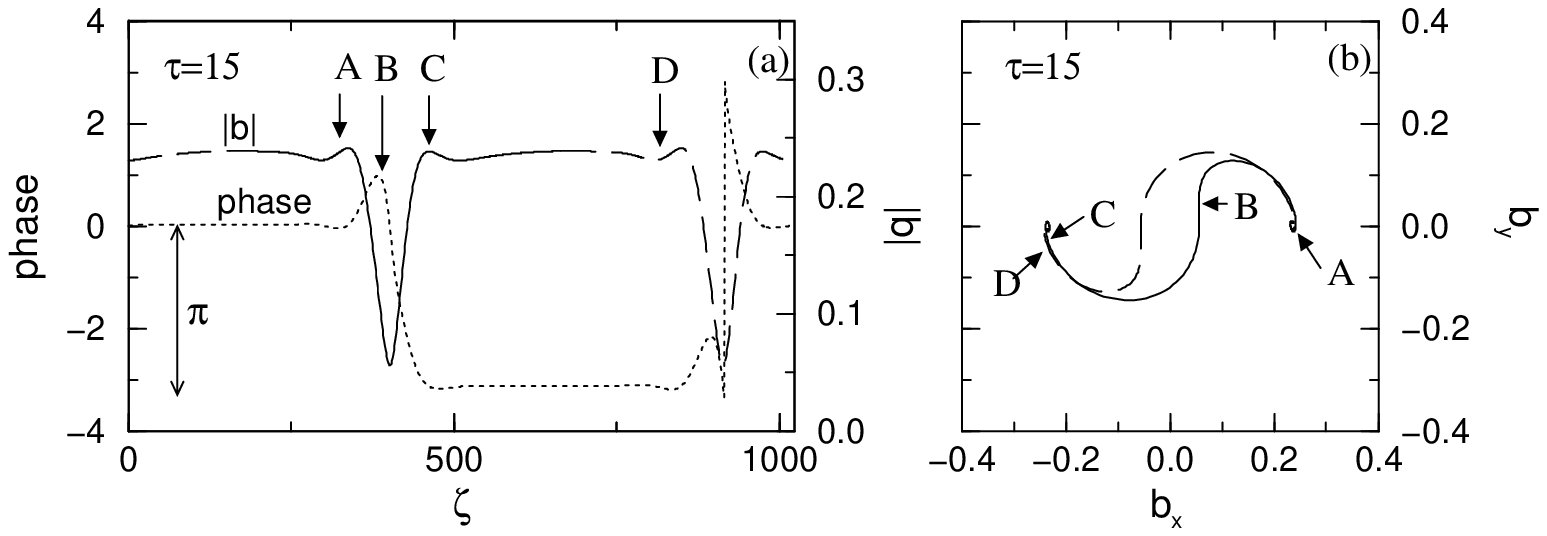}\vspace*{-8in} Fig.~3\vskip0.5in
\hspace*{-1in}\psfig{file=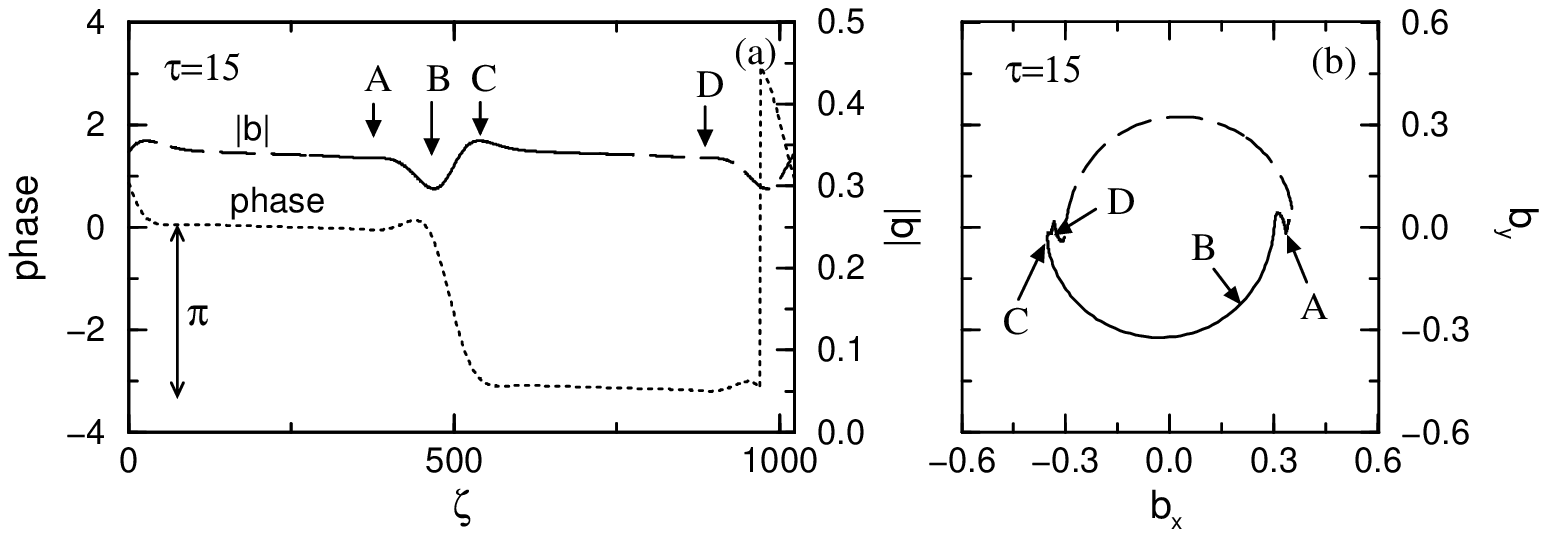}\vspace*{-8in} Fig.~4\vskip0.5in
\hspace*{-1in}\psfig{file=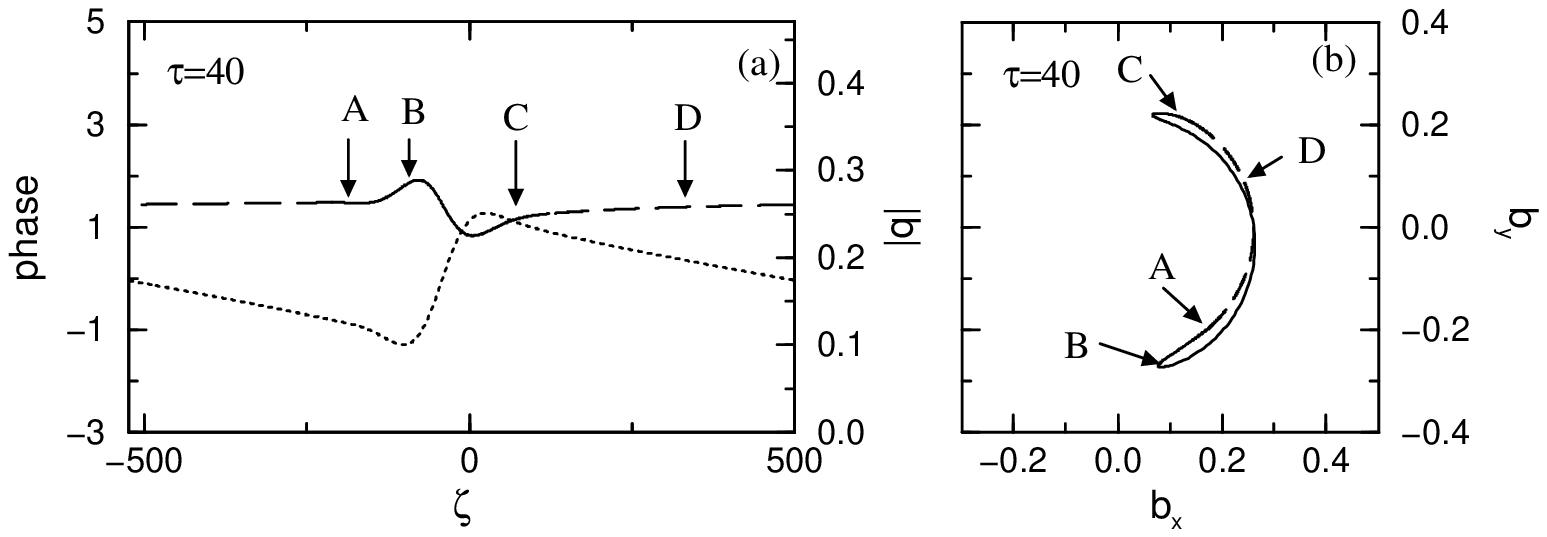}\vspace*{-8in} Fig.~5\vskip0.5in
\thispagestyle{empty}
\end{document}